\documentclass[]{aa}
\usepackage[varg]{txfonts}
\usepackage{graphicx}
\usepackage{natbib}
\bibpunct{(}{)}{;}{a}{}{,} 
\graphicspath{{hdimmagini/}}

\title{Exoplanet atmospheres with GIANO}
\subtitle{II. Detection of molecular absorption in the dayside spectrum of HD 102195b}

\author{G.~Guilluy		\inst{\ref{unito},\ref{oato}}
	\and A.~Sozzetti	\inst{\ref{oato}}
	\and M.~Brogi	\inst{\ref{warwick}, \ref{oato}, \ref{warwick2}}
	\and A.S.~Bonomo	\inst{\ref{oato}}
	\and P.~Giacobbe	\inst{\ref{oato}}
	\and R.~Claudi \inst{\ref{padova}}
	\and S.~Benatti \inst{\ref{padova}}
}
\institute{Dipartimento di Fisica, Universit\`{a} degli Studi di Torino, via Pietro Giuria 1, I-10125 Torino, Italy \label{unito}
	\and INAF – Osservatorio Astrofisico di Torino, Via Osservatorio 20, I-10025 Pino Torinese, Italy \label{oato}
	\and Department of Physics, University of Warwick, Coventry CV4 7AL, UK \label{warwick}
	\and Centre for Exoplanets and Habitability, University of Warwick, Gibbet Hill Road, Coventry CV47AL, UK \label{warwick2}
	\and INAF – Osservatorio Astronomico di Padova, Vicolo dell’Osservatorio 5, 35122, Padova, Italy \label{padova}
}

\date{Received <date> / Accepted <date>}
\abstract{The study of exoplanetary atmospheres is key to understand the differences between their physical, chemical and dynamical processes. Up to now, the bulk of atmospheric characterization analyzes has been conducted on transiting planets. On a number of sufficiently bright targets, high-resolution spectroscopy (HRS) has also been successfully tested for non-transiting planets mainly by using spectrographs mounted on 8- and 10-m class ground-based telescopes.}
{The aim of this analysis is to study the dayside of the non-transiting planet HD 102195b using the GIANO spectrograph mounted at the Telescopio Nazionale Galileo (TNG), and thereby demonstrate the feasibility of atmospheric characterization measurements, and in particular of molecular detection, for non-transiting planets with the HRS technique using 4-m class telescopes.  }
{Our data analysis technique exploits the fact that the Doppler-shifted planetary signal changes on the order of many km s$^{-1}$ during the observations, in contrast with the telluric absorption which is stationary in wavelength. This allows us to effectively remove the contamination from telluric lines in the GIANO spectra while preserving the features of the planetary spectrum. The emission signal from HD 102195b's atmosphere is then extracted by cross-correlating the residual GIANO spectra with models of the planetary atmosphere.}
{We detect molecular absorption from water vapor at the 4.4$\sigma$ level of statistical significance. We also find convincing evidence for the presence of methane, which is detected at the 4.1$\sigma$ level. This is the first detection of methane obtained with the HRS technique. The two molecules are detected with a combined significance of 5.3$\sigma$, at a semi-amplitude of the planet radial velocity $K_{\rm P}=128 \pm 6$ km s$^{-1}$. We estimate a planet true mass of $M_{\rm P}=0.46 \pm 0.03$ \(\textup{M}_J\) and constrain the orbital inclination in the range 72.5$^{\circ}$<$i$<84.79$^{\circ}$ (1$\sigma$).	Our analysis indicates a non-inverted atmosphere for HD 102195b. This is expected given the relatively low temperature of the planet, inefficient to keep TiO/VO in gas phase. Moreover, a comparison with theoretical model expectations corroborates our detection of methane, and a cursory confrontation with chemical model predictions published in the literature, suggests that the detected methane and water signatures could be consistent with a low C/O ratio for HD 102195b. Finally, being HD 102195 1-3 magnitudes fainter in the K-band than non-transiting systems studied until now with 8-m telescopes, our study allows us to open the doors of the atmospheric characterization to a larger sample of exoplanets.}
{}

\begin{document}  
	\maketitle
	
	\section{Introduction}
	Ground-based and space surveys have uncovered an astonishing diversity in the physical properties of exoplanets (e.g. \citealt{2017AFulton} and references therein). Nevertheless, there are still many unanswered questions relative for example to their true nature, including their internal composition and its possible dependence on the formation history. Exoplanet atmospheres provide a means to directly address these still open questions. 
	
	Atmospheric characterization measurements have been particularly successfull for the class of transiting exoplanets.
	The spectra of transiting planets are obtained by studying how the total flux (star+planet) varies as a function of wavelength, when the planet passes in front (transmission spectrum) or behind (emission spectrum) its host star along our line-of-sight. 
	The probability that a planet transits is given by $P_{transit} \sim R_\star/a$, where $a$ is the orbital separation between the star and the planet, and $ R_\star$ is the star’s radius. For close-in hot Jupiter exoplanets, this probability is on the order of $\sim 10 \%$ or less. Unfortunately, this type of exoplanets are generally far away from us; the planets around nearby bright stars typically do not transit. For this reason, we need to devise a method capable of studying the atmospheres of non-transiting planets around bright stars in the solar neighborhood.
	
	In the past, many studies have been carried out trying to characterize the atmosphere of non-transiting planets with the high-resolution spectroscopy (HRS) technique. \citet{Cameron1999} first reported a probable detection of starlight from $\tau$ Boötis b. This was followed by other works \citep{Charbonneau1999, Wiedemann2001, Leigh2003, Cameron2004, Lucas2009, Rodler2010}, some very useful since they provide meaningful upper limits that confirmed low albedos for hot Jupiters. But unfortunately, searches for atmospheric compounds, in these earlier works, did not result in robust  detections. A turning point came with \citet{Brogi2012} and \citet{Rodler2012}. By using a cross-correlation technique similar to that developed by \citet{Snellen2010} in the case of the transiting planet HD 209458b, they succeeded in detecting a molecule (carbon monoxide) in the atmosphere of a non-transiting planet: the hot Jupiter $\tau$ Bo\"otis b.
	
	At high spectral resolution, molecular features are resolved into a dense forest of individual lines in a pattern that is unique for a given molecule \citep{Birkby2018}, so molecular species (H$_2$O, CO, CO$_2$, CH$_4$, etc.) can be robustly identified by line matching with planetary models. In addition to providing information on atmospheric composition, the HRS technique can also provide an estimate of the true mass and the orbital inclination of non-transiting planets by breaking the $M_{\rm P}\sin {i}$ degeneracy intrinsic to the radial velocity (RV) measurements. In fact, HRS allows to detect orbital velocity of a planet in the form of a Doppler-shifted spectrum. In this way, the planetary mass can be determined using the Newton's law of gravity, treating the star+planet system as a double-line spectroscopic binary.

	Atmospheric characterization studies for non-transiting planets at HRS have been carried out using spectrographs mounted on the biggest telescopes on the ground, such as CRIRES at VLT \citep{Rodler2012, Brogi2013, Brogi2014, Snellen2014, Schwarz2016, Birkby2017} or NIRSPEC at Keck \citep{Lockwood2014, Piskorz2016, Piskorz2017}. In this paper, we show that also smaller telescopes, if equipped with performing high-resolution spectrographs, can be employed for molecular detections in the dayside spectra of non transiting planets. Our group has worked with the Telescopio Nazionale Galileo (TNG) and we have recently demonstrated that, despite its 3.6-m mirror, it can be successfully employed to characterize the atmosphere of transiting planets around bright host stars \citep{Brogi2018}. The TNG is so performant because it is endowed with the high resolution ($ R\sim$ 50,000) near-infrared spectrograph GIANO \citep{Oliva2006, Origlia2014}. GIANO can simultaneously cover a wide spectral range (0.95-2.45) $\mu$m in each individual exposure. It provides a 25-fold increase in spectral range compared to CRIRES pre-upgrade at VLT.
	Furthermore, this large spectral range allows GIANO to satisfy the request in terms of signal-to-noise for securing solid detections in the NIR.
	
	In this work, we report spectroscopic observations of the dayside of the non-transiting hot giant planet HD 102195b with GIANO, to probe its atmospheric composition and to estimate the planet's true mass and the inclination angle of the orbital plane.
	
	The paper is organized as follows. In Section 2, we describe the system HD 102195. The observations and the standard calibration are discussed in Section 3, while in Section 4 we report how we extract the planetary signal. The results of the analysis are presented in Section 5. The summary and discussion are provided in Section 6.

	\section{The system HD 102195}\label{system}
	
	HD 102195 is a mildly active G8V relative bright ($V=8.06 \pm 0.01$ mag, $J=6.629 \pm 0.024$ mag, $H=6.268 \pm 0.026$ mag, $K=6.151 \pm 0.018$ mag) dwarf \citep{Ge2006, Meloetal2007}, that may be part of the Local Association \citep{Ge2006}.  
	We have redetermined the stellar parameters by using the Yonsei-Yale evolutionary tracks \citep{Demarque2004} with constraints on the stellar luminosity $L=0.485 \pm 0.014\,\rm L_\odot$ from the Gaia DR2 parallax \citep{Gaia2016, Gaia2018}, and the host star effective temperature ($T_{\rm eff}=5291 \pm 34$ K) and metallicity ($ \rm [Fe/H]=0.05 \pm 0.05$ dex) as derived by \citet{Meloetal2007} from the analysis of HARPS spectra. We found $M_\star=0.890 \pm 0.020\, \rm M_\odot$ and $R_\star=0.837 \pm 0.017\, \rm R_\odot$, which are consistent with the values given both in \citet{Ge2006} and \citet{Meloetal2007}.
	
	The discovery of a giant planet around HD 102195 was announced by \citet{Ge2006}. It was discovered with the RV measurements  gathered with the Exoplanet Tracker instrument, a dispersed fixed-delay interferometer, mounted at the Kitt Peak National Observatory (KPNO) telescope. Additional radial velocities were also gathered with the HARPS spectrograph at the 3.6~m ESO telescope (La Silla) by \citet{Meloetal2007}.
	Since HD 102195b is a non-transiting planet only Doppler studies have been possible, from which we have an estimate of its minimum mass $M_{\rm P}\sin {i}=0.488 \pm 0.015$ \(\textup{M}_J\) \citep{Ge2006}, while its orbital inclination $i$ is undetermined.
	In this work, we use the HRS technique to determine planetary true mass and system inclination.
	\subsection[Orbital Ephemeris]{Orbital Ephemeris}\label{ephemeris}  
	To refine the orbital ephemeris, we modeled both the HET/High-Resolution-Spectrograph and ESO/HARPS radial velocities with a Keplerian orbit by employing a differential evolution Markov chain Monte Carlo Bayesian technique \citep{Eastmanetal2013} with the same implementation as in \citet{Bonomoetal2015}. We found the improved orbital period P and inferior conjunction time to be $P=4.11390 \pm 0.00072$ d and $T_{\rm {c}}=3897.068 \pm 0.061$ $\rm BJD_{\rm TDB}$, respectively. We did not include the radial velocities obtained at the KNPO telescopes because they do not yield any significant improvement in the orbital solution, their formal uncertainties being approximately one order of magnitude larger than those of the HRS and HARPS measurements (see Table 1 in \citealt{Ge2006}). The orbital eccentricity $e=0.056 \pm 0.036$ is consistent with zero (e.g., \citealt{LucySweeney1971, Bonomoetal2017}).

	\section{Observation and standard calibration}
	We observed HD 102195 with the near-infrared high-resolution echelle spectrograph GIANO mounted at the Nasmyth-A focus of the TNG. GIANO operated in this configuration at the TNG in the March 2015 - August 2016 period. GIANO was fiber-fed: the light reached it through two optical fibers (each with an angular diameter of 1") at a fixed distance of 3". 
	GIANO covered 4 spectral bands in the NIR (Y, J, H, K) divided into 47 orders. 
	The spectra were imaged on a $2048 \times 2048$ detector.
	Observations of science targets were performed with the nodding on fiber acquisition mode: every target and sky were taken in pairs and alternatively acquired on nodding A and B, respectively, for an optimal subtraction of the detector noise and background.
	Each order was thus divided into two different tracks: orderA and orderB.  Furthermore, the presence of an additional slit divided each position (orderA, orderB) into two tracks: A1, A2, B1, B2.
	
\begin{figure}
		\resizebox{\hsize}{!}{\includegraphics{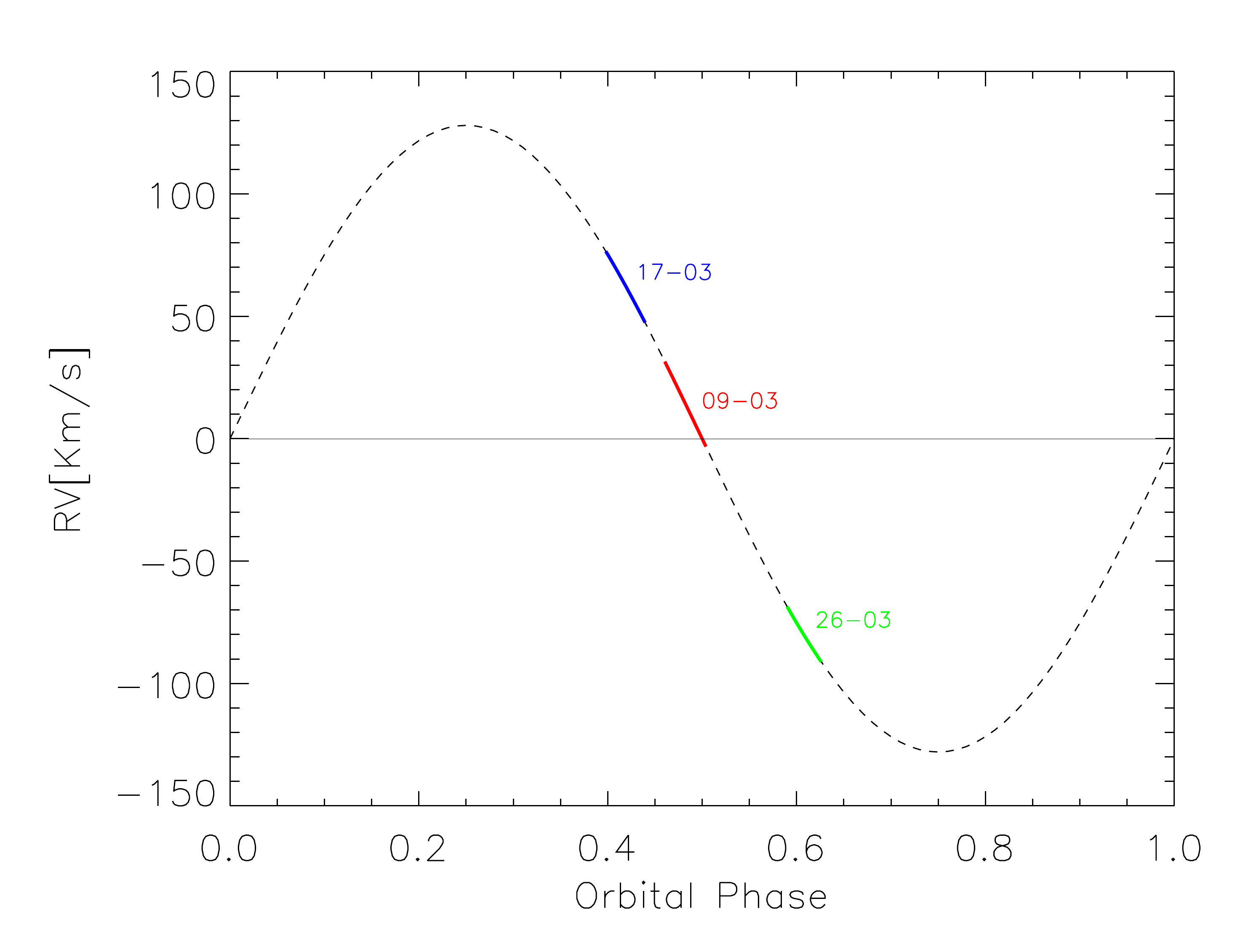}}
		\caption{Radial velocity curve for HD 102195b. The planet has been observed during the nights of 9 (red), 17 (blue), 26 (green) March 2016, covering three ranges of planetary orbital phases 0.46-0.50, 0.40-0.44 and 0.59-0.63. During these phase intervals, the planet orbital motion changes by tens of km s$^{-1}$, while the telluric/stellar contamination are stationary signals in wavelength. 
		}
		\label{orbita_2}
	\end{figure}

	GIANO observations of HD 102195 were carried out during the nights of 9, 17, 26 March 2016, collecting a total of 48, 50, and 46 spectra, respectively, each with an exposure time of 200 s. During these nights of observation, three ranges of planetary orbital phase were targeted: $\varphi=0.46-0.50$, $\varphi=0.40-0.44$ and $ \varphi=0.59-0.63$ (see Figure \ref{orbita_2}). In these phase intervals, the planet is near the superior conjunction where it reaches the maximum rate of change of velocity. The radial component of the planet's orbital motion changes by tens of km s$^{-1}$, which greatly helps in disentangling the Doppler-shifted planet spectrum from the stationary signals in wavelength due to telluric contamination and to stellar spectrum.

	\subsection[Extraction of 1-D spectra]{Extraction of 1-D spectra}\label{extra}
	For these observations, we have written our own analysis pipeline in IDL, including the calibration of the data and the extraction of the one-dimensional spectra.
	A master flat-field is produced by subtracting dark images, taken with the same exposure time, and averaging the difference images in time.
	The master flat-field is used to identify the orders' position. 
	The position of the apertures is then refined by fitting each spectral track with four Gaussians per order simultaneously (two sliced images and two nodding positions). The location of the orders is given by the average of the four Gaussian centroids. This operation is carried out for 32 equidistant positions along the dispersion direction. The sampled values are then fitted with a parabolic polynomial on the 2048 pixels of the dispersion direction.
	Since the spectral tracks are curved they are rectified via spline interpolation. 
	
	At this step, it is possible to work with A and B images (i.e. images taken respectively in the nodding positions A and B). 
	Each image is then corrected for the bad pixels by assigning zero flux to the damaged pixels.
	In order to remove the thermal background and sky emission lines, it is appropriate to work with difference (A$-$B) images.
	The one-dimensional spectra are then extracted via   optimal extraction \citep{Horne1986}. The B spectra need to be multiplied by $-1$ to correct for the initial subtraction of the B images. The optimal extraction does not preserve the fluxes. To overcome this problem every spectrum of each aperture must be scaled so that its average is equal to that of the spectrum obtained with rectangular extraction.

	Even if a correction before extracting the spectra has been performed, there could be residual bad pixels to be corrected. Order by order and image by image the remaining bad pixels are identified by applying a sigma clipping to each spectrum (after subtracting out the median-filtered version of the same spectrum). 
	For the two nodding positions, numerical masks are built to account for the bad pixels present in each image and in each order.

	\subsection[Aligning the spectra]{Aligning the spectra}\label{align}
	
	Each spectrum is then divided by its blaze function. Order by order the blaze function is created fitting a 4-order polynomial to the spectrum obtained from the master flat with the same extraction procedure described in Section \ref{extra}.
	
	The scientific analysis exploits the fact that the Earth’s atmosphere transmission spectrum and, to a first approximation, the stellar spectrum are stationary signals in wavelength (i.e. they fall always in the same pixels) while during each of the three nights of observations the planetary radial velocity is significantly Doppler-shifted and changes on the order of tens km s$^{-1}$ (i.e., it moves across many pixels during the observing sequence). 
	For this reason, we have to correct the fact that the wavelength solution of GIANO is not stable during hours of observation and this causes small shifts in wavelength. We remove this effect by aligning all the observed spectra between them.

	For each night the spectral alignment is then performed by computing the cross-correlation between a reference spectrum (taken as the average over time) and each spectrum order by order and for a lag vector of 401 elements between -2 pixels and +2 pixels.
	
\begin{figure*}
		\centering
		\includegraphics[width=17cm]{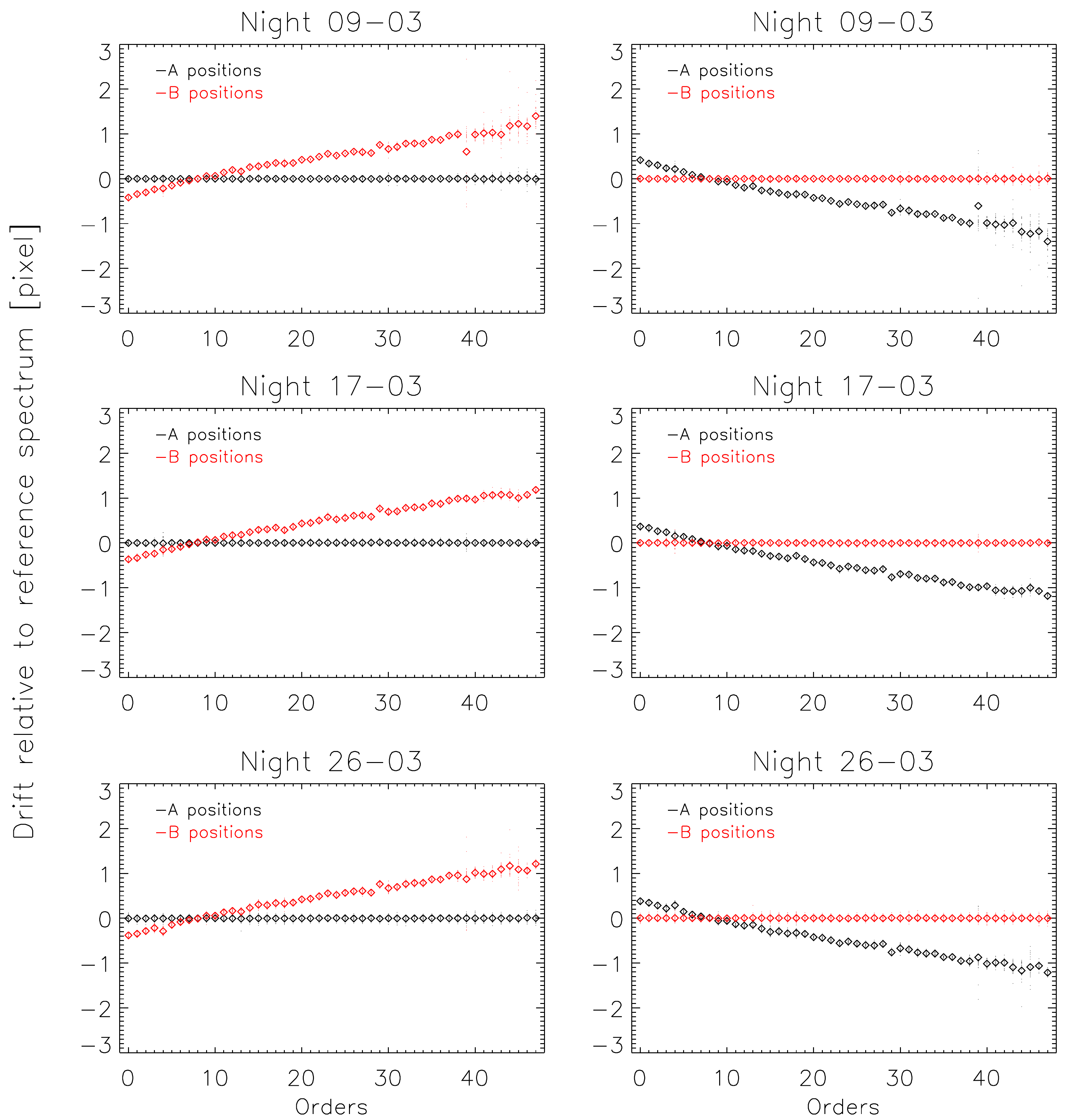}
		\caption{For each order, we plot the average shift (diamonds) over time (dots). The drift relative to the reference spectrum is expressed in pixels (1 pixel equals to 2.7 km s$^{-1}$).		
			In the left panels, for the three nights, we have taken as reference spectrum the averaged over the spectra taken in B position  (in red) while in the right ones the average over the spectra taken in A position (in black). 
			The average shifts over time of the spectra acquired with a nodding position are shifted from those taken in the other one in a linear way as a function of spectral orders. The shifts amplitude doesn't depend on the night considered.}
		\label{allinemento}
	\end{figure*}

	The fractional lag corresponding to the maximum cross-correlation is adopted as the shift of each spectrum compared to the reference one.
	Figure \ref{allinemento} shows that there is a linear trend as a function of orders between the dispersion solution for the A nodding position (in black) and for the B one (in red). This effect had already been noticed in the GIANO data collected for the HD 189733b atmospheric characterization study \citep{Brogi2018}. Order by order, the trend is removed by re-aligning each spectrum to the reference one by spline interpolating it based on the calculated shifts.

	\subsection[Wavelength solution]{Wavelength solution}\label{Wave}
	The absorption spectrum of the Earth's atmosphere (telluric spectrum) in the NIR provides an excellent wavelength calibration source. For each order, we visually match the positions of a set of telluric lines in the time-averaged observed spectrum and in a high-resolution Earth's atmospheric transmission spectrum generated via the ESO Sky Model Calculator\footnote{\url{https://www.eso.org/observing/etc/bin/gen/form?INS.MODE=swspectr+INS.NAME=SKYCALC}}.
	The telluric lines are chosen to be isolated and not saturated. The line centroids are then estimated by fitting each line with a Gaussian profile. The centroid positions are computed in pixel for the averaged spectrum, and in wavelength for the corresponding lines in the template spectrum. 
	A first (pixel, wavelength) relation is obtained by fitting the centroids positions with a fourth-order polynomial.
	We then use this first solution to refine the centroid positions; we compute a super-sampling of each line via spline interpolation at 1/20 of a pixel and we record the fractional pixel and the corresponding wavelength at which the flux reaches a minimum. We re-fit this new (pixel, wavelength) relation with a fourth-order polynomial and we assume the fit as the wavelength solution.
	
	Unfortunately this type of calibration based on the telluric lines is not possible for all the orders. For those showing a few or no telluric lines, the wavelength calibration steps are performed by working with the stellar lines instead of the telluric lines,  and by using as reference spectrum a Phoenix stellar spectrum \citep{Allard2011}, calculated for a star with properties closely matching those of HD 102195. 
	
\begin{figure}
		\resizebox{\hsize}{!}{\includegraphics{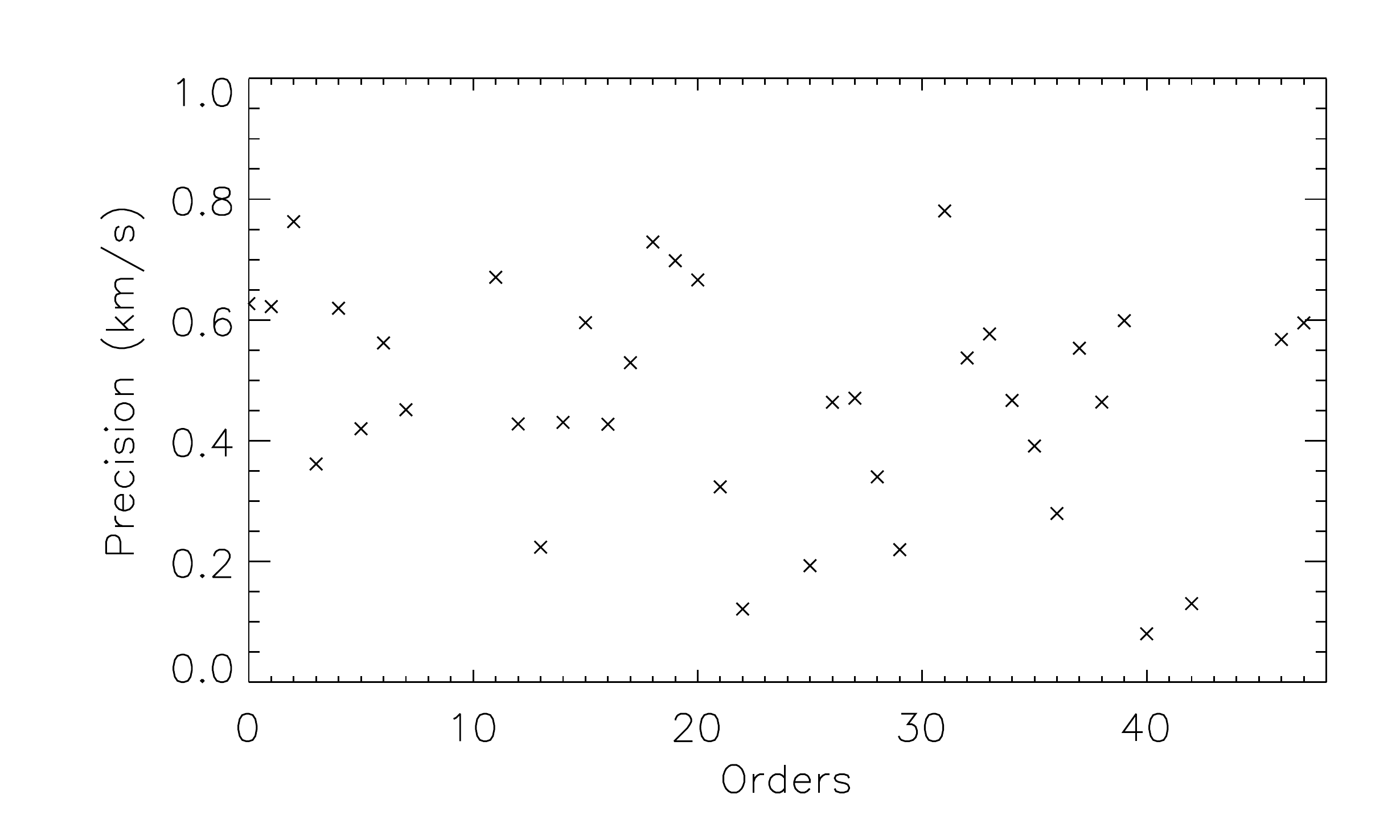}}
		\caption{Precision of the wavelength calibration for the orders of GIANO considered in the analysis of HD 102195b spectra.}
		\label{calib}
	\end{figure}

	Since there are spectral orders where the Earth's atmosphere is very opaque (i.e. they do not have enough flux or unsaturated spectral lines), and others in which there is a mismatch between modeled and observed spectral lines, we are not able to calibrate all the spectral regions, and we have to exclude orders 8, 9, 10, 23, 24, 30, 41, 43, 44, 45 from our analysis.
	For the remaining orders, the precision of our calibration is estimated by computing the standard deviation of the residuals. We achieve a residual scatter per line well below 1 km s$^{-1}$, approximately 1/3 of a pixel (see Figure \ref{calib}). 

	\section[Extraction of the planetary signal]{Extraction of the planetary signal}
	In ground-based observations, the planetary signal is hidden under telluric contamination.
	Standard techniques to remove telluric lines typically use the spectrum of a telluric standard star observed before or after each science spectrum. However, because we want to observe how the planet changes its velocity during time and gain as much signal-to-noise as possible, it is not advisable to interrupt the series of observations to take telluric standards. Moreover, telluric standards are observed in regions of the sky different from the studied target and with different airmasses. Since in the NIR the atmosphere varies on timescales of minutes, the correction with telluric standards would not reach the level of precision required by our analysis.
	There is an alternative approach, a key point of the HRS method, that allows us to disentangle the planet and the telluric spectrum and therefore to remove the telluric contamination \citep{Birkby2018}. 
	The planetary signal during the three nights of observation is significantly Doppler-shifted and changes on the order of tens km s$^{-1}$, while Earth’s atmosphere transmission spectrum is a stationary signal in wavelength. The telluric spectrum can, therefore, be removed by modeling the flux of each spectral channel in time. For this analysis, we follow the approach of \citet{Brogi2018}.
	\subsection[Removing the telluric lines]{Removing the telluric lines}  \label{remove}
\begin{figure}
		\includegraphics[width=\linewidth]{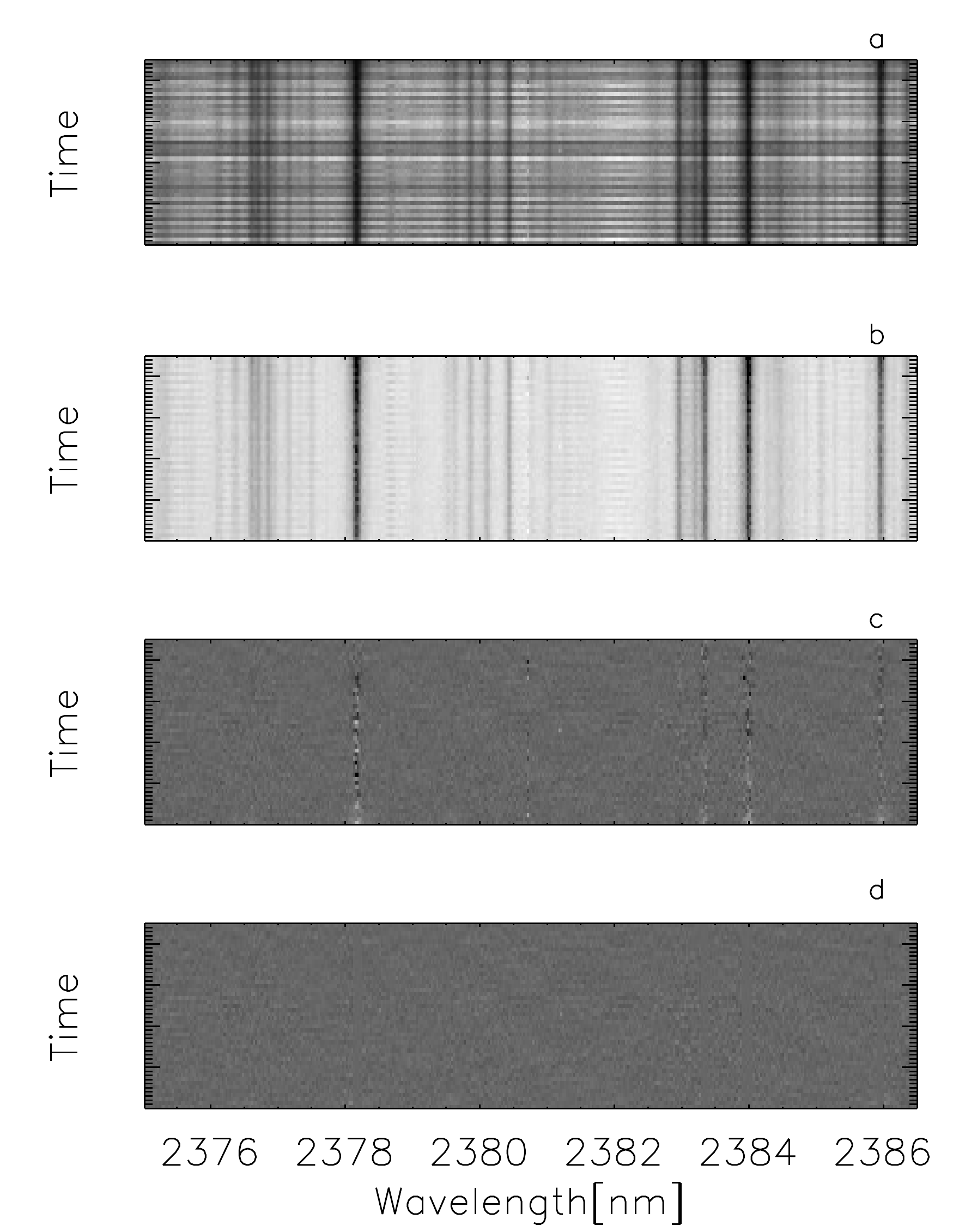}
		\caption{An example of our data reduction chain to remove the telluric contamination. a) The extracted spectra. b) Data in logarithm space after the throughput correction. c) After the airmass correction. d) Finally GIANO's residuals after telluric correction.}
		\label{operazioni}
	\end{figure}

	We remove the telluric contamination as follows (see Figure \ref{operazioni}). 
	For each night we consider every order separately as a matrix with wavelength (pixels, 2048 channels) on the $x$-axis and time (frame number, $n_{f}$) on the $y$-axis. We prefer treating separately the A and B spectra to take into account possible variations between the two nodding positions.  So, we work with $n_{ord}$ (orders positions remained after wavelength calibration) $\times$ 2 (nodding position) matrices, each one sized 2048 $\times$ $n_f$.
	We first apply the masks built in Section \ref{extra} and we update them including all the pixels that present a flux less than 0.1 counts. We perform this operation because we work with flux in logarithmic space (since the depth of telluric lines depends on the exponential of airmass) and in this way we avoid infinite values. We mask also residual damaged pixels, previously identified in Section \ref{extra}.
	The masks are re-updated whenever additional bad pixels are identified.
	Once the natural logarithm of each matrix has been taken, to remove the instrumental throughput (due to pointing, seeing, and sky transparency) we subtract the median of the 300 brightest pixels from each spectrum.
	
	Since the GIANO detector is divided into 4 sensors each one sized 1024 $\times$ 1024, every order is so split into two sensors. To correct for the differences between the right and the left part of each order, pixels 0 through 1023 along the $x$ axis are separated from pixels 1024 through 2047. For the three nights of observation, we thus work with $n_{ord}$ (positions of the orders) $\times$ 2 (nodding position) $\times$ 2 (detector's quadrants) matrices sized 1024 $\times$ $n_f$. From every spectrum of these matrices a linear fit between the spectrum itself and a time-average spectrum is subtracted out. In this way, we eliminate possible discontinuities between spectra acquired with different nodding position or with different detector quadrants. The linear fit is performed using the IDL routine \textit{ladfit.pro}, which fits the paired data (xi, yi) with a linear model, y = A + Bx, using a robust, outlier resistant least absolute deviation method.

	The most dominant feature in the spectra remains the telluric absorption, which has a fixed pattern in wavelength but has a depth which varies with time (due to changes in airmass during the night of observation). For this reason, we remove the telluric contamination by modeling, for every spectral channel, $\lambda_i$, the airmass changes with time. For every order, we firstly merge the matrices corresponding to the right and left part of the GIANO quadrant and we subtract from each spectrum a second order fit performed between the observed flux (in logarithmic space) and the airmass. Some residuals are still present at the position of some of the strongest absorption lines due to varying conditions in Earth's atmosphere during the night. This is expected because the amount of water vapor changes, generally not following the airmass, and it is not distributed homogeneously in the atmosphere, but rather confined to the Earth's troposphere. We correct for this effect by measuring the flux changes in few of the deepest H$_2$O lines over time, and we use these to correct the rest of the matrix with a column-by-column linear regression \citep{Brogi2014}.

	After the telluric removal, the bad pixels/low counts masks are applied again assigning zero flux to these pixels.
	To eliminate residual strong outliers a sigma-clipping with a threshold of eight times the standard deviation of each matrix is applied. 
	In the end, the matrices are returned to linear space.

	\subsection[Atmospheric Models]{Atmospheric Models} \label{modelli}
	\begin{figure}
		\includegraphics[width=\linewidth]{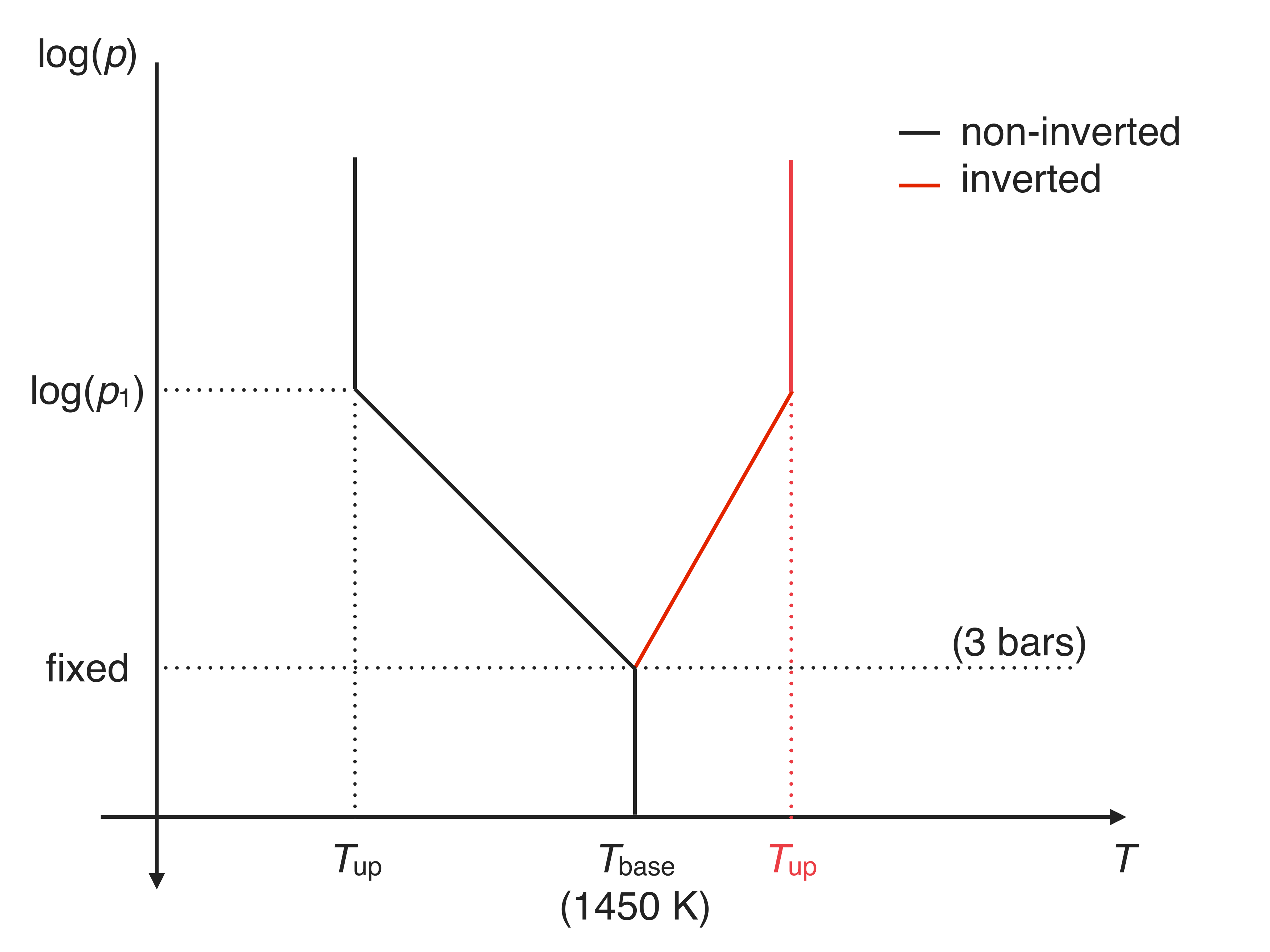}
		\caption{	Schematic representation of the parametrisation used in our models for the planet temperature-pressure profile. The atmosphere is composed of two isothermal layers (at temperatures $T_\mathrm{base}$ and $T_\mathrm{up}$) separated by a layer with constant lapse rate $d\log(p)/dt$. While $T_\mathrm{base}$ is fixed to 1450 K and the base pressure is fixed to 3 bars, the other quantities are varied in a small grid, as explained in Section \ref{modelli}.}
		\label{TP_profile}
	\end{figure}
	
	We compute models for the atmosphere of HD 102195b by using the line-by-line plane-parallel radiative transfer code of \citet{dekok14}. The planet atmosphere is divided into 50 layers equally spaced in logarithm of pressure ($\log{p}$), and the corresponding temperature is interpolated from a prescribed temperature-pressure profile given as input (see Figure \ref{TP_profile}). Opacities from H$_2$-H$_2$ collision-induced absorption \citep{bory01, bory02} are incorporated. CO, H$_2$O, and CH$_4$ are added as trace species, with line position, intensity, and air-broadening parameters taken from HITEMP2010 \citep{rothman2010} for CO and H$_2$O, and from HITRAN2012 for CH$_4$ \citep{rothman2012}. Due to the overwhelming amount of transitions, the line list for water vapour is peeled to keep the strongest 10,000 lines every 0.042 $\mu$m in wavelength, resulting in a total of 380,000 H$_2$O lines. The line lists for CO and CH$_4$ do not require peeling, and contain 40,000 and 46,300 lines, respectively, in the range of these GIANO models (0.91-2.50 $\mu$m). Being extrapolated from low-temperature measurements, the HITRAN methane list is known to be incomplete, however it contains the majority of the strong lines shorter than 1.6 $\mu$m and longer than 1.7 $\mu$m, where most of the GIANO data is sensitive to \citep{Yurchenko14}. In order to keep the modeling computationally feasible, we calculate two sets of models:
	\begin{enumerate}
		\item In the first set, water, carbon monoxide, and methane are added as single trace species. The temperature-pressure profile is bounded by two isothermal zones, one at temperature $T_\mathrm{base} = 1450$ K for $p>3$ bar, and the other at $T=T_\mathrm{up}$ for pressures $p<p_1$, where $\log(p_1) = [-4, -3, -2]$ bar and $T_\mathrm{up} = [600, 850, 1650]$ K. Volume Mixing Ratios for the species range between 10$^{-6}$ and 10$^{-3}$ in steps of one decade. We thus compute 36 models per species.
		\item In the second set, we mix the two species detected singularly (water and methane) and we further simplify the temperature-pressure profile by removing the thermal inversion case ($T_\mathrm{up} = 1650$ K), which is only producing anti-correlation from the previous set of models (see Section \ref{Moleculesidentification}). We thus produce a set of 54 models.
	\end{enumerate}
	
	\subsection[Extraction of the signal by Cross Correlation]{Extraction of the signal by Cross Correlation} \label{CC}
	With the telluric lines removed as above, we are left with residuals (shown in the lower panel of Figure \ref{operazioni}). Since the individual planet lines have very low SNR ($SNR_{line}$ $<< $ 1) the planet signal is deeply hidden in the noise. However, there are thousands of molecular lines in the GIANO spectral range, so we can combine their signals to attempt a detection of the planet signature. 
	
	The spectral lines can be co-added by cross-correlating the residual GIANO data with a model spectrum appropriate for the planet atmosphere, in this way observation by observation we evaluate how well the model matches the residual spectra.

	The cross-correlation is computed on a fixed grid of radial velocity lags between -225 and 225 km s$^{-1}$, in steps of 2.7 km s$^{-1}$ (167 values), which matches the pixel scale for most of the GIANO orders. We must be careful to choose the step size, because, if it is taken smaller than the velocity resolution of the detector's pixel, we risk an oversampling of the data \citep{Birkby2018}.
	The cross-correlation is calculated for every order and for every integration separately. Each spectrum is cross-correlated with the model, Doppler shifted for each RV value, and convolved with the instrument profile of GIANO (a Gaussian with FWHM ~5.4 km s$^{-1}$). 
	For every night the output of the cross-correlation is a matrix with dimensions 167$\times$$n_f$$\times$$n_{ord}$.
	
	Each cross-correlation function (CCF) gives an estimate of how well the model matches the data for each RV step; a positive value indicates correlation, while negative values denote anti-correlation. This can arise if we cross-correlate emission lines with models in absorption or vice-versa. 
	
	The planetary signal is not visible yet. To maximize it, we co-add all orders obtaining three matrices $n_f$x167 one for each night.
	
\begin{figure}
		\includegraphics[width=\linewidth]{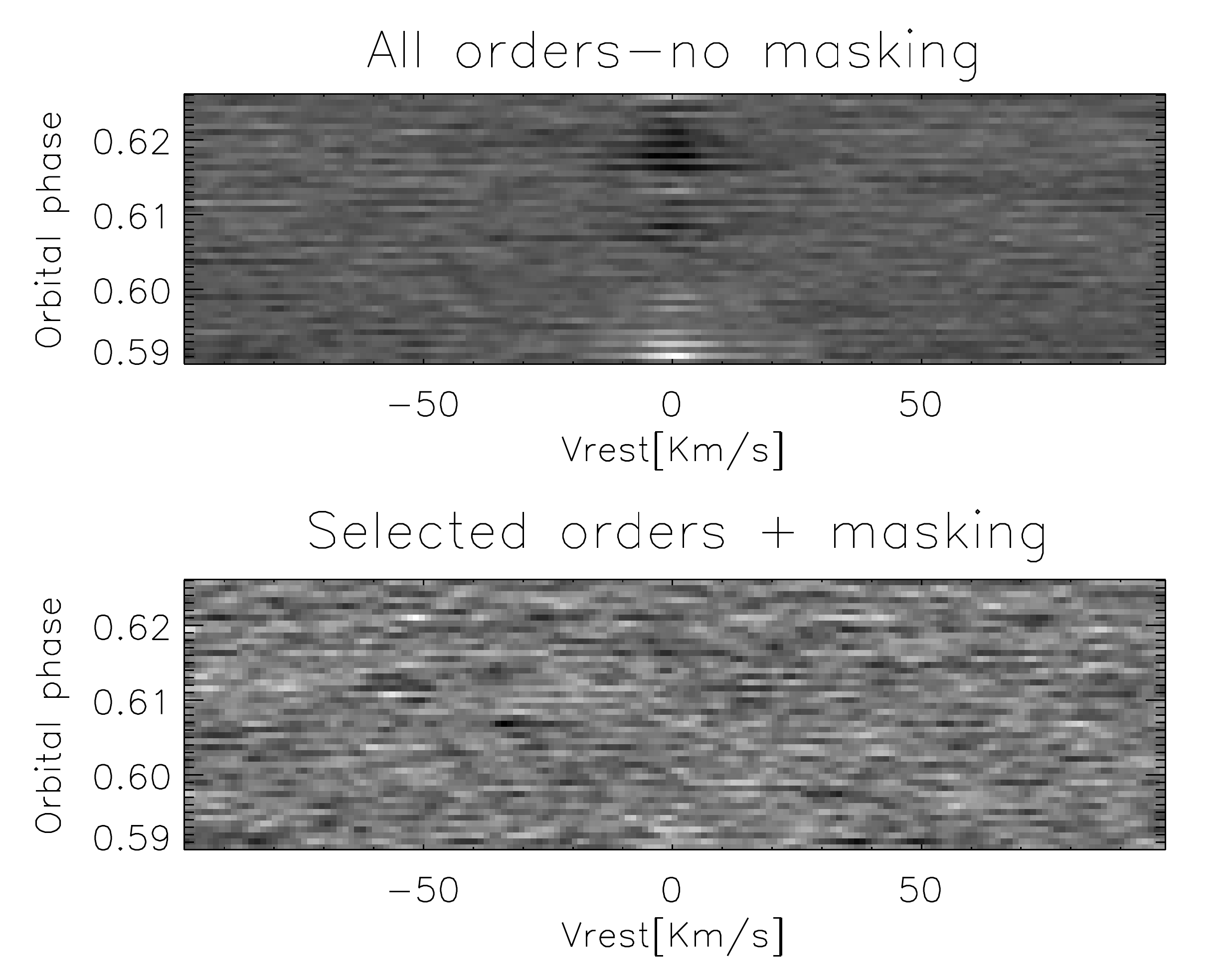}
		\caption{Co-added matrix of CCFs for the night of March 26.
			In the upper panel at zero radial velocity, telluric residual are visible. We remove them applying a mask to the data before the cross-correlation. The result of this operation is shown in the lower panel.}
		\label{residui}
	\end{figure}

	The top panel of Figure \ref{residui} illustrates the coadded matrix of CCFs for the night of March 26. It shows clear telluric residuals, which are at zero velocity in the telluric frame.
	We therefore apply the masking of \citet{Brogi2018} prior to cross-correlation by assigning 0 value to those pixels corresponding to telluric lines deeper than 5-30 $\%$, where this threshold varies order by order. This is a procedure that is to be done iteratively by visually analyzing the cross-correlation of each order until telluric residuals disappear. 
	Despite this masking, in some orders we continue to see some cross-correlation noise that does not resemble the signature of residual telluric lines (likely affected by e.g. modal noise). We therefore discarded these orders from the analysis. In this way we remain with the following groups of orders  [0, 1, 2, 3,32 ,33 ,37 ,38 ,40 ,41], [0, 2, 7, 17, 18, 19, 25, 27, 29, 31, 32, 33, 34, 38, 40, 41], [1, 2, 3, 4, 12, 14, 16, 19, 27, 29, 32, 33, 40, 41], for the nights of 09, 17, and 26 March, respectively.
	Finally, we proceed to co-add the cross-correlation matrices of the three nights on time. 
	Since every CCF is taken at a different planetary phase and so at a different radial velocity, we must shift all the CCFs in the planet rest frame.
	Doppler shifting the planet requires the knowledge of its radial velocity ($V_{rv}$), the systemic velocity of the star-planet system with respect to the Earth, for which we adopt the Gaia DR2 value $V_{sys}=1.85 \pm 0.15$ km s$^{-1}$ \citep{Gaia2016, Gaia2018}, and the observer's velocity induced by the Earth's motion around the Sun ($V_{obs}$) (i.e. the barycentric velocity). Assuming a circular orbit and neglecting cross-terms in velocity, the total planetary velocity $V_{\rm P}(t)$ is given by:  
	\begin{equation}
	V_{\rm P}(t)=V_{rv}+V_{sys}+V_{obs}
	\end{equation}
	
	\noindent{In the above equation, $V_{rv}=K_{\rm P}\,\sin{[2\pi\,\varphi(t)]}$, where $\varphi(t)$ is the planetary phase and  $K_{\rm P}$ is the semi-amplitude of the planet radial velocity. Since HD 102195b is a non-transiting planet, $\sin{i} $ is undetermined, so we do not know $K_{\rm P}$. We assume a range of possible $K_{\rm P}$ (0$\le K_{\rm P} \le $ 170) km s$^{-1}$, which matches all the possible orbital inclination of the system, plus a small interval of unphysical  values ($\sin$ $i > 1$) as a sanity check \citep{Brogi2014}.} The phase $\varphi(t)$ at a time $t$, is obtained as the fractional part of
	\begin{equation}
	\varphi(t)=\frac{t-T_{\rm {c}}}{P}
	\end{equation}
	where  $P$ is the orbital period and $T_{\rm {c}}$ is the the time of inferior conjunction (see Section \ref{ephemeris}). Uncertainties in $T_{\rm {c}}$ and $P$ translate into a 1$\sigma$ uncertainty of $\Delta \varphi=$0.15 on the orbital phases.

	For each value of $K_{\rm P}$, we align the CCFs in the planet rest frame via linear interpolation (so that they are centered around $V_{\rm P}(t)$), and we sum them up in phase, thus combining the signal from all of the lines present in the spectra. 
	In this way we obtain the total planetary cross-correlation signal as a function of the rest frame velocity $V_{rest}$ and the planetary semi-amplitude $K_{\rm P}$. 
	If the planetary signal is detected, we will measure a peak in the cross-correlation map at $V_{rest} \sim 0$. Significant deviations far from $V_{rest} \sim 0$ indicates that the signal does not have a planetary origin.
	
	\section{Results}
	\subsection[Molecules identification]{Molecules identification} \label{Moleculesidentification}
	
\begin{figure*} 
		\centering
		\includegraphics[width=17cm]{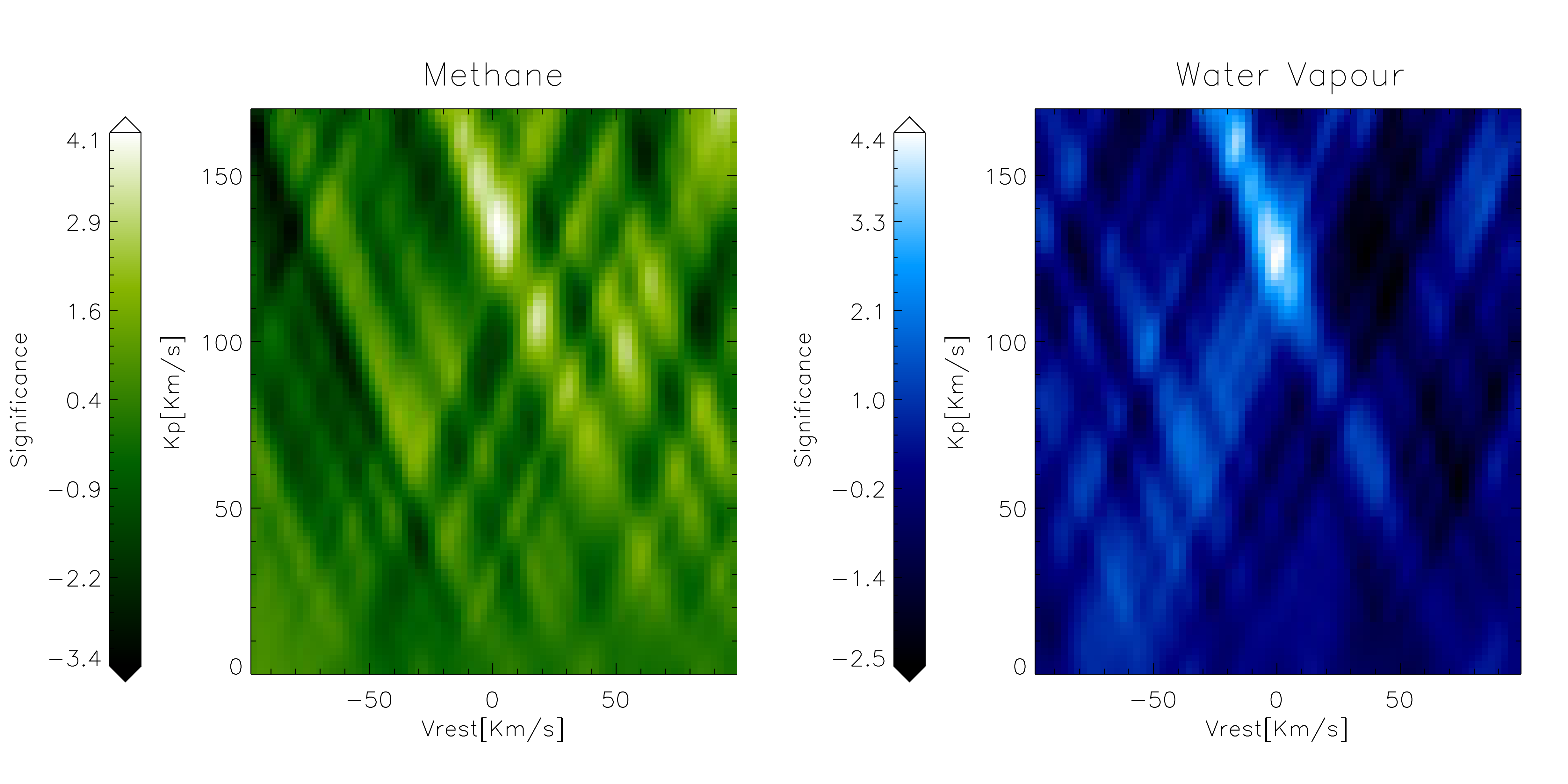}
		\caption{Total planetary cross-correlation signal, obtained by combining the three nights, as a function of the rest frame velocity $V_{rest}$ and the planetary semi-amplitude $K_{\rm P}$. The cross-correlation has been performed using models containing only one molecule: CH$_4$ (left panel) or H$_2$O (right panel). The significance is calculated, using the Welch t-test, as described in Section \ref{significance}}
		\label{molecole}
	\end{figure*}

	Figure \ref{molecole} shows the results of the cross-correlation obtained from models containing only one molecule, i.e. water vapor or methane.
	H$_2$O and CH$_4$ are detected with a significance respectively of 4.4$\sigma$ and 4.1$\sigma$. The significance level is estimated through the Welch t-test as explained in Section \ref{significance}.  
	
	CO is not detected in these data at significance above 3$\sigma$. Although this might seem counter-intuitive compared to past CRIRES studies in which this molecule produced the strongest detections, one ought to consider the different type of sensitivity of CRIRES compared to GIANO. The latter instrument, being mounted at a smaller telescope and having less resolving power, can only compete with CRIRES by significantly increasing the spectral range to include multiple NIR bands. However, this benefit only applies to the detection of molecular species broadly absorbing in the NIR, particularly water. CO, on the other hand, only absorbs in part of the K band, and more weakly in part of the H band. At the signal to noise of these GIANO observations, there are not enough CO lines to build up a significant detection through cross-correlation. We verified indeed that injecting CO-only model spectra in the data and attempting to retrieve their signal produces no firm detections, with peak values of significance below 3.0. For these reasons, we will not consider CO for the remainder of the analysis.
	
\begin{figure*} 
		\centering
		\includegraphics[width=17cm]{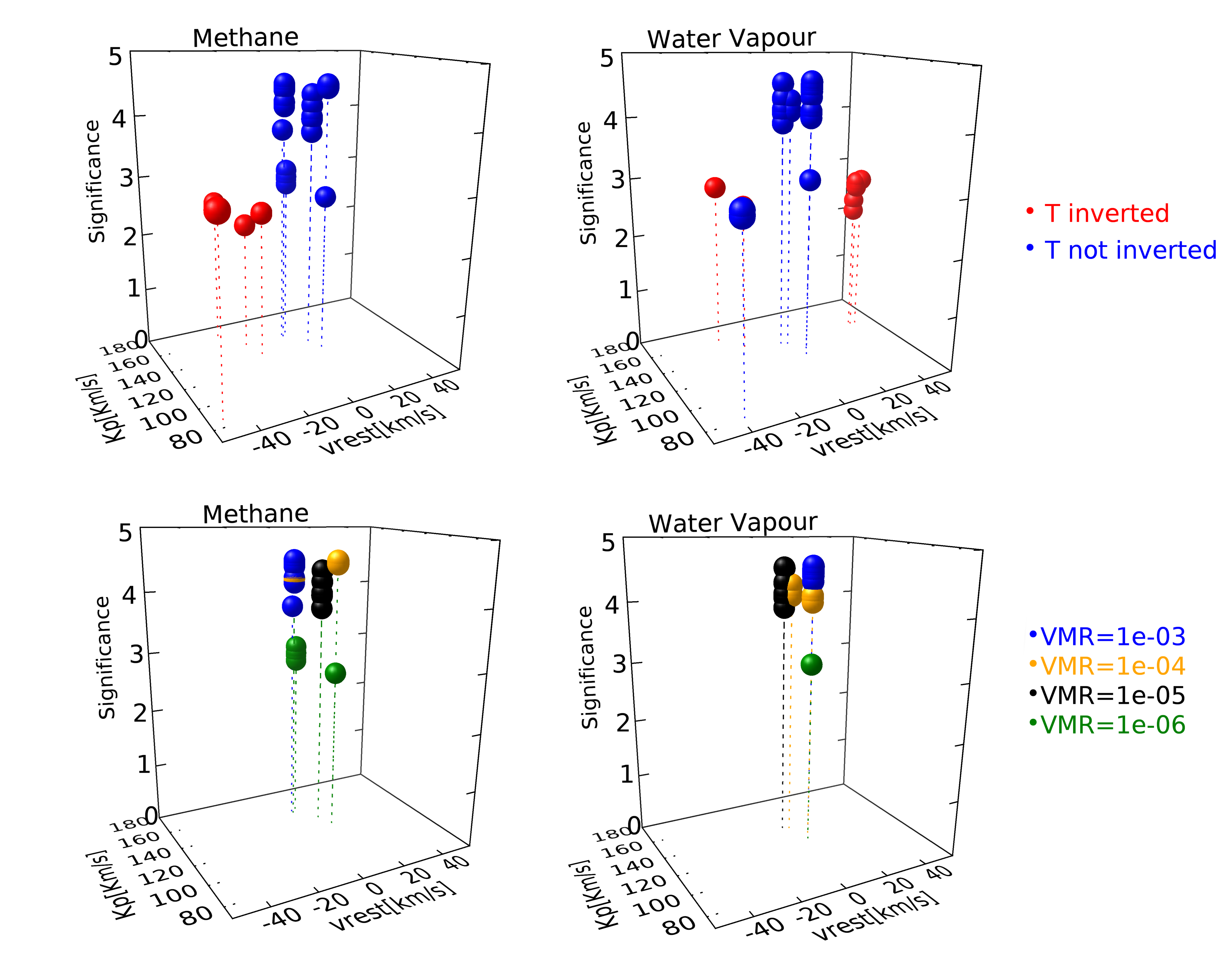}
		\caption{Significance, as a function of the rest frame velocity $V_{rest}$ and the planetary semi-amplitude $K_{\rm P}$, for the grid of models containing only water vapour (right panels) or methane (left panels). From the upper panels, we see that model with inverted T/P profiles (red bullets) are not good descriptors of the observations. Indeed they have a low significance and peak far away from $V_{rest} \sim 0$, where we would expect a signal of planetary origin. Then, if we restrict our analysis to the models which peak near $V_{rest}$ $\sim$ 0 we find that three families of models provide better agreement than others both for water vapour ($10^{-5}<$VMR(H$_2$O)$<10^{-3}$) and for methane ($10^{-5}<$VMR(CH$_4$)$<10^{-3}$). The different bullets shown in the bottom panels correspond to the different lapse rates used.}
		\label{modelli_fig}
	\end{figure*}

\begin{figure*} 
		\centering
		\includegraphics[width=17cm]{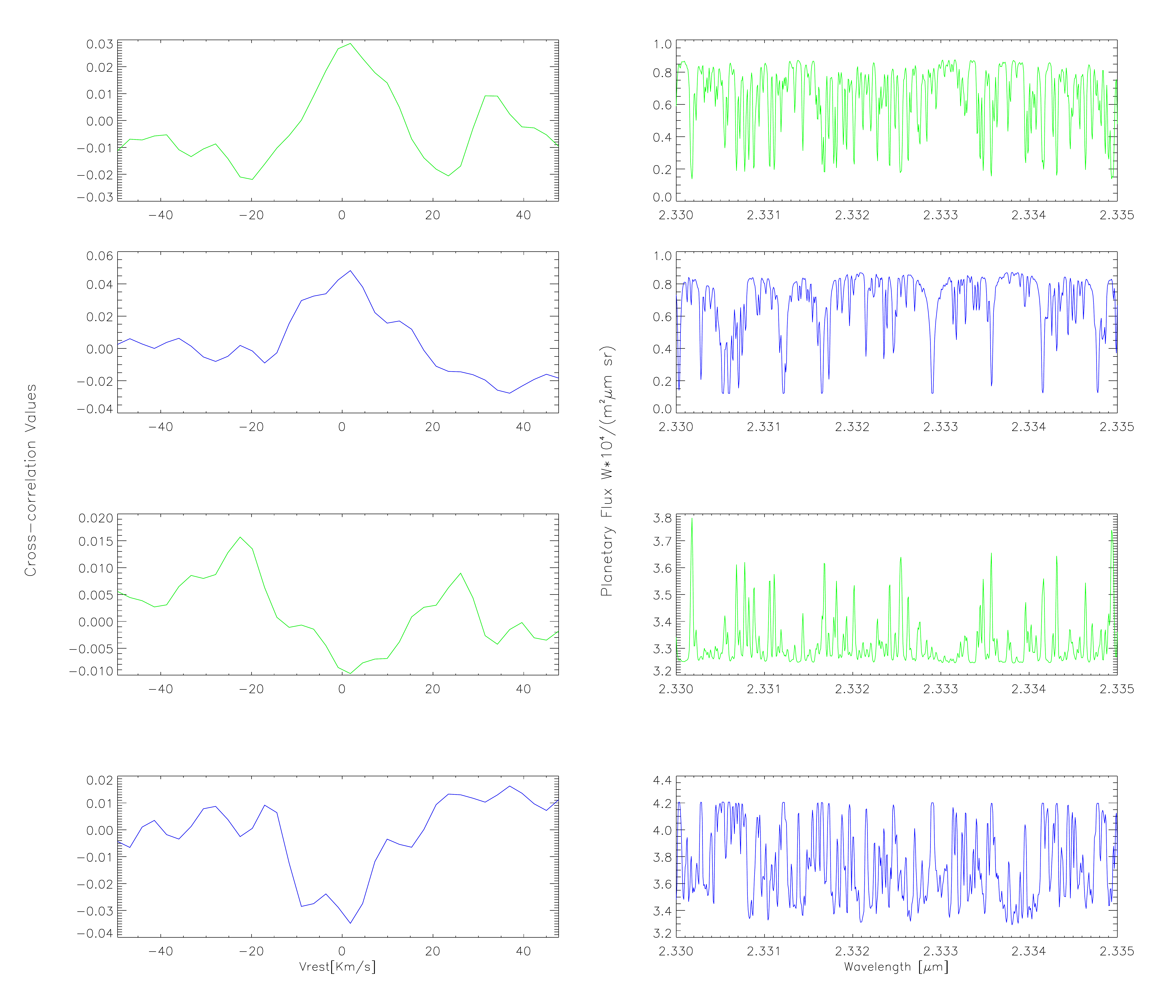}
		\caption{Total cross-correlation values (left panels) in the neighborhood of $V_{rest} \sim 0$, and theoretical models (right panels) they correspond to. We use the best models containing methane-only (in green, first panels), and water vapour-only (in blue, second panels), and two models with temperature inversion with methane-only, VMR $\sim 10^{-5} $, (in green, third panels), and water vapour-only, VMR $\sim 10^{-3} $, (in blue, fourth panels). Since the emission lines in the models with temperature inversion correlated negatively with the absorption lines in the observed planetary spectrum, unlike the absorption lines in the non-inverted models, we can infer a not-inverted temperature profile for HD 102195b.}
		\label{cc_tuttimod_}
	\end{figure*}

	We choose the best-matching dayside model by requiring the highest significance of the cross-correlation signal obtained from the combination of the three nights. We can immediately rule out an inverted temperature profile for HD 102195b. Indeed, as the upper panels of Figure \ref{modelli_fig} shows, when we cross-correlate with grid models with a temperature inversion (i.e., T2 = 1650 K), they do not show significant cross-correlation and they peak at $V_{rest}$ $\neq$ 0. Moreover, the emission lines of these models tend to negatively correlate with the absorption lines in the observed planetary spectrum (see for example Figure \ref{cc_tuttimod_}, and the discussion in \citealt{Birkby2017,Schwarz2015}). For these reasons we conclude that inverted models in our grid do not correctly represent our data.
	
	After this first screening, if we exclude from our analysis the models which peak far from $V_{rest}$ $\sim$ 0, and we consider the water vapor molecule alone (the panel at the bottom right of Figure \ref{modelli_fig}), we find that three families of models appear to behave better than others, i.e. those with Volume Mixing Ratio (VMR) $ \sim 10^{-3}$, (VMR) $ \sim 10^{-4}$, and (VMR) $ \sim 10^{-5}$. These are selected requiring at the same time proximity to $V_{rest}$ $\sim$ 0, high significance, and $K_{\rm P}$ similar or less than the orbital velocity of the planet $V_{\rm P}$ (see Section \ref{inclination}). If we treat the models containing methane-only (the panel at the bottom left of Figure \ref{modelli_fig}), the models with $10^{-5}<$VMR(CH$_4$)$<10^{-3}$ seem to provided the best agreement.  
	
	From the single-molecule analysis, after testing the full  set of models described in Section \ref{modelli}, we find that the cross-correlation signal comes from H$_2$O and CH$_4$. The VMRs appear to be between $10^{-3}$ and $10^{-5}$ for both molecules, perhaps with a (slight) preference towards $10^{-5}$ for CH$_4$ and towards $10^{-3}$ for H$_2$O. For both molecules, the preferred lapse rate brackets the range 150 K/dex<dT/dlog$_{10}$(${p}$)<425 K/dex.

	Finally, the total cross-correlation signal, obtained from models containing both methane and water vapor, appears to be maximized  when we consider a mixed model with VMR(H$_2$O) $\sim 10^{-3}$ and VMR(CH$_4$) $\sim 10^{-5}$.
	
\begin{figure} 
		\resizebox{\hsize}{!}{\includegraphics{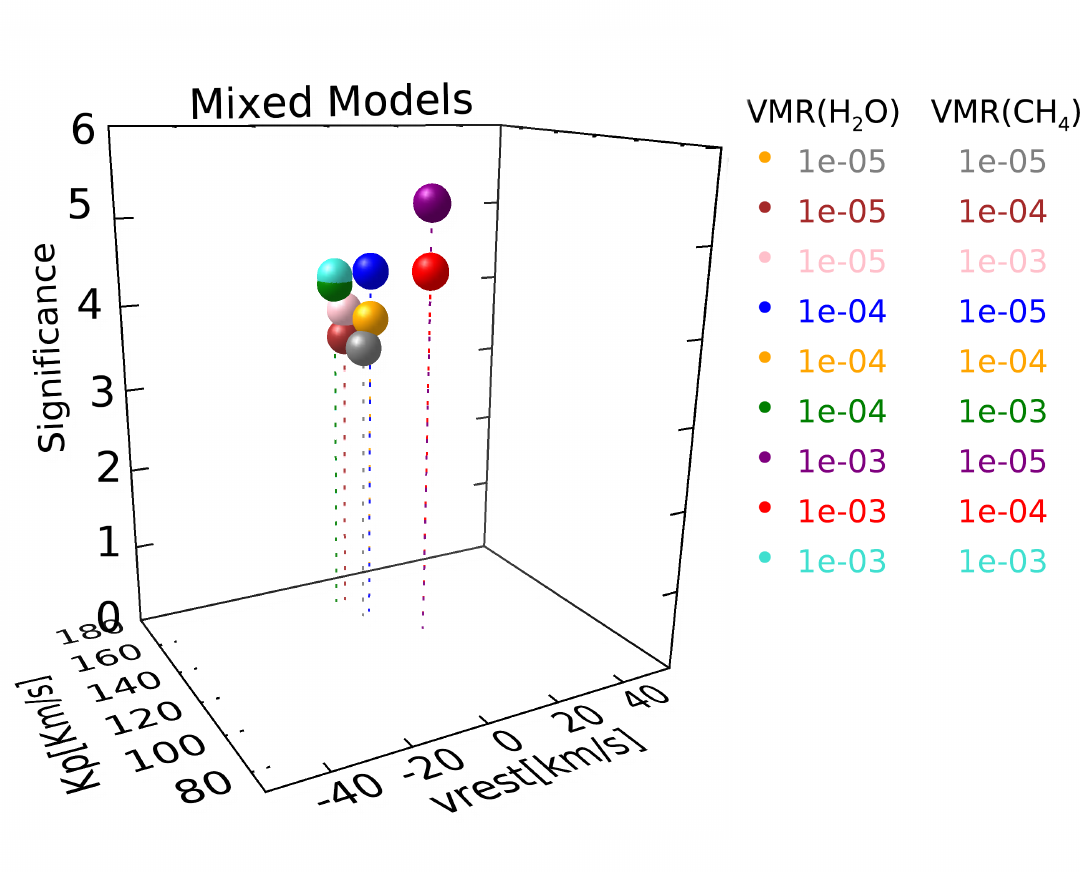}}
		\caption{Significance, as a function of the rest frame velocity $V_{rest}$ and the planetary semi-amplitude $K_{\rm P}$, by assuming a dT/dlog$_{10}$(p) $\sim$ 425 K/dex. From this mixed models analysis, we found that the model with VMR(H$_2$O) $\sim 10^{-3}$ and  VMR(CH$_2$4) $\sim 10^{-5}$ seems to be preferred marginally over the others. }
		\label{tutti_fig}
	\end{figure}

	If we consider the nights taken singularly and their combination, within the small group of models (VMR(H$_2$O) $\sim 10^{-3}$,  VMR(CH$_4$) $\sim 10^{-5}$, 150 K/dex<dT/dlog$_{10}$(p)<425 K/dex) the one with dT/dlog$_{10}$(p) $\sim$ 425 K/dex is marginally preferred. To better illustrate our VMRs choise, we plot in Figure \ref{tutti_fig} the different significance level of the mixed models, assuming a dT/dlog$_{10}$(p) $\sim$ 425 K/dex. 
	We do note that multiple models with different physical model parameters yield detections with similar statistical significance, and therefore the best model is preferred only  marginally over the others.
	
	\subsection[Planetary Mass and System Inclination]{Planetary Mass and System Inclination} \label{inclination}

	The total cross-correlation signal, obtained from the best model (VMR(H$_2$O)=$10^{-3}$,  VMR(CH$_4$)=$10^{-5}$, dT/dlog$_{10}$(p) $\sim$ 425 K/dex) by combining the three nights of observation, is shown in the rigth panel of Figure \ref{ttest}. The cross-correlation peak is detected at the orbital semi-amplitude of $K_{\rm P}\sim 128 \pm 6$ km s$^{-1}$. A phase shift of $\Delta \varphi=0.012$ needs to be applied in order to match the signal at zero $V_{rest}$. This shift is well within the 1$\sigma$ uncertainty range of  $\Delta \varphi=0.15$.

\begin{figure}[tb]
		\includegraphics[width=\linewidth]{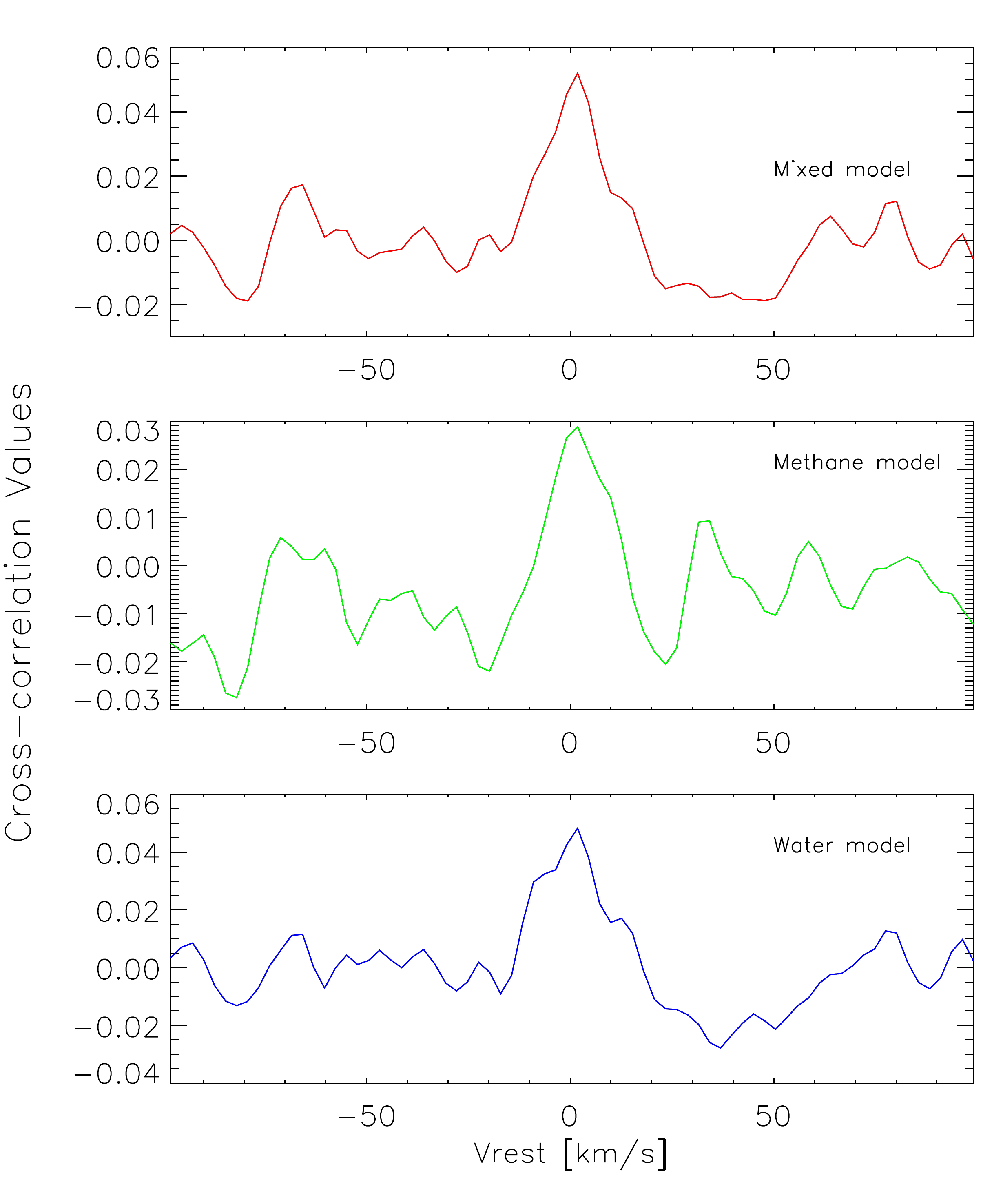}
		\caption{Total cross-correlation signal as a function of $V_{rest}$ from all spectra taken during the three nights combined assuming a planet orbital velocity corresponding to the cross-correlation peak. We use our best water-only, methane-only, and mixed models, in the lower, middle, and upper panel, respectively.}
		\label{CC128}
	\end{figure}

	The upper panel of figure \ref{CC128} shows the one-dimensional cross-correlation function for the mixed model, that best matches the data, assuming a planet orbital velocity corresponding to the cross-correlation peak. For completeness, the middle and lower panels show the one-dimensional CCF computed for methane-only, and water-only models, respectively.

	The measurement of $K_{\rm P}$ allows us to treat the star-planet system as a double-lined spectroscopic
	binary. We can therefore derive the true mass of HD 102195b, through the velocity and mass ratios $M_S/M_{\rm P}=K_{\rm P}/K_S$. 
	Given the stellar RV semi-amplitude $K_S=63.4\pm2.0$  m s$^{-1}$ \citep{Ge2006}, its mass  $M_S=0.89\pm 0.02$ \(\textup{M}_\odot\), and our planetary semi-amplitude $K_{\rm P} = 128 \pm 6 $ km s$^{-1}$, we obtain a first estimate of the planetary mass of $M_{\rm P}=0.46 \pm 0.03$ \(\textup{M}_J\).  The major source of uncertainty in the derivation of the planetary mass comes from the error on $K_{\rm P}$ which represents about 47$\%$ of the indetermination in the calculation of $M_{\rm P}$.
	
	The final step is to derive the orbital inclination. Indeed by using the Kepler's Third Law, the orbital velocity of the planet is found to be $V_{\rm P}=128.0 \pm 1.0$ km s$^{-1}$, and under the assumption of a circular orbit $\sin{i}=K_{\rm P}/V_{\rm P}$, therefore allowing us to determine $\sin{i}=1.00 \pm 0.05$. This is translated into 1$\sigma$, 2$\sigma$, 3$\sigma$ lower limits on the orbital inclination of 72.5$^o$, 65.0$^o$ and 59.2$^o$ respectively. The non detection of the planet in previous photometric monitoring \citep{Ge2006} places an upper limit on the system inclination of $i_{max}=84.79^o \pm 0.09^o$.
	We note that the resulting error on the assumed radial velocity of the planet is 1.0 km s$^{-1}$, and being on the order of 1/3 of the GIANO pixels, it does not significantly affect the co-addition of the planetary signal.

	\subsection [The Significance of the Cross-correlation Detection]{ The Significance of the Cross-correlation Detection} \label{significance}
\begin{figure*} 
		\centering
		\includegraphics[width=17cm]{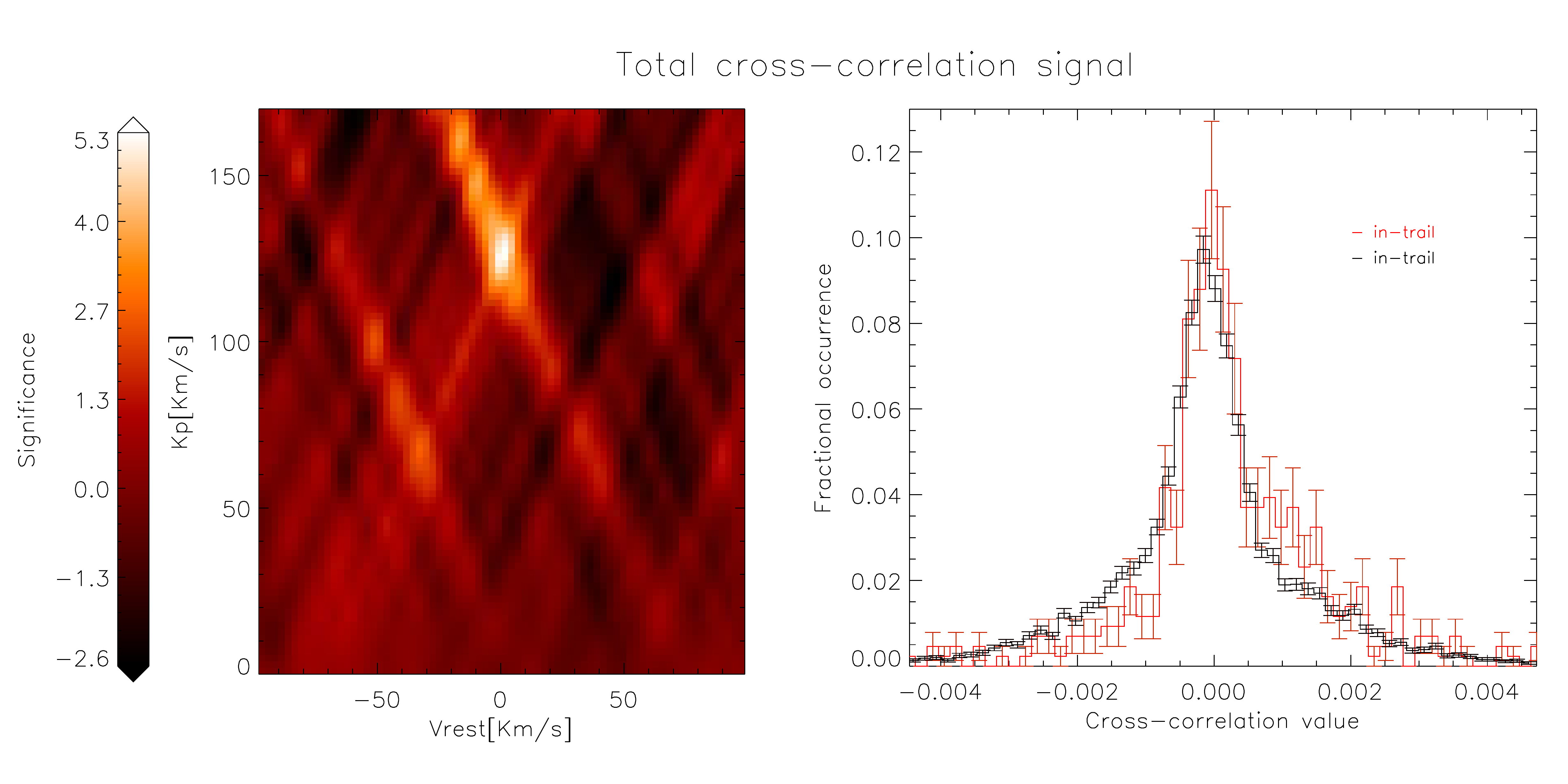}
		\caption{The histogram in the right panel shows the difference between the aligned CCF distributions `in-trail' (red-line) and `out-of-trail' (black-line) for the derived $K_{\rm P}$. The two distributions are normalized and the error bars corresponds to the square-root of the bin occurrences. The in-trail distribution is centered to higher values than the out-of-trail distribution with a 5.3 $\sigma$ confidence level. The results of the Welch t-test performed for the same range of $v_{rest}$ and $K_{\rm P}$ as before is plotted in the left panel.}
		\label{ttest}
	\end{figure*}

	To secure the exoplanet's spectroscopic detection we have to determine the significance level of the signal. Following \citet{Brogi2012, Brogi2013, Brogi2014, Brogi2018}, we compare the `in-trail' values, in the two dimensional aligned CCFs' matrix, to those `out-of-trail'. The right of  Figure \ref{ttest} shows that the in-trail distribution, for the derived $K_{\rm P}$ (i.e. those within 4 pixels from the planetary RV curve) is shifted toward higher values than the out-of-trail one (i.e. those 10 pixels away from the planet radial velocity). In order to quantify this, a Welch t-test on the two samples is generated \citep{Welch1947}, with the null hypothesis H$_0$ that, if they are drawn from the same parent distribution, they have the same mean. The corresponding $t$ value is converted into a significance at which H$_0$ is rejected, based on the results of the test, that is 5.3$\sigma$.
	The left panel of Figure \ref{ttest} shows also the test performed for the same range of $v_{rest}$ and $K_{\rm P}$ used before.  Except for the peak of the cross-correlation map, no other region of the parameter space shows significant cross-correlation signal.

	We may think that atmospheric models containing water vapor and methane would give a cross-correlation signal with a significance that corresponds to the quadrature sum of that those obtained from models with H$_2$O or CH$_4$ only, in this case we would obtain a significance of $\sim 6 \sigma$. 
	As we have seen above, this is not the case, as we have obtained a lower combined signal at 5.3$\sigma$. However, this lower significance can be explained by noting that the two cross-correlation signals are not completely independent, as water and methane lines can overlap at the instrumental resolution of our observations, reducing the total amount of strong lines contributing to the cross-correlation function.
	\section{Summary and discussion} 
	
	We have studied the atmosphere of the non-transiting gas giant HD 102195b at high spectral resolution using the near-infrared spectrograph GIANO. We have observed the system during three nights in March 2016, during which the planet was near the superior conjunction. This helped us to distinguish between the telluric contamination and the planetary signal. In this way, we have detected absorption from water vapor and from methane in the planetary dayside hemisphere.
	
	After correcting the telluric lines, in order to detect the planetary signal, we have cross-correlated the GIANO residual data with a wide grid of models for the planetary atmosphere. We have obtained that the best model has minimum molecular abundances of VMR(H$_2$O)=$10^{-3}$, VMR(CH$_4$)=$10^{-5}$ and a steep lapse rate of dT/dlog$_{10}$(p) $\sim$ 425 K/dex. 
	Our results are in general agreement with the two-dimensional (2D) classification scheme suggested by \citet{Madhusudhan2012}. Indeed, for temperatures close to that of HD 102195b, they predict a difference of $\sim 10^{-2} $ between the two VMR. In particular, a cursory comparison with chemical model predictions published in that paper allows us to infer that a VMR difference of $\sim 10^{-2} $ is compatible with a possibly oxygen-rich atmosphere for HD 102195b, with a C$/$O value qualitatively comparable to the solar one.

	Moreover, our analysis indicates that models with a thermal inversion are disfavoured. This result is in line with the predictions of the theory. In fact, the host star being active, $\log R'_{HK}=-4.5$ \citep{Meloetal2007}, according to \citet{Knutson2010}, we should not have expected a temperature inversion, because the UV flux of active stars might destroy the molecular species responsible for this mechanism. Also, a non-inverted atmosphere is expected given the relatively low temperature of the planet; indeed we believe that the condensation of TiO / VO plays an important role, potentially able to remove these species in the gas phase from the atmosphere.

	The results of the cross-correlation analysis presented here also allowed us to derive the semi-amplitude of the planetary radial velocity ($K_{\rm P} = 128 \pm 6 $ km s$^{-1}$), which could be translated in a value for the true planetary mass ($M_{\rm P}=0.46 \pm 0.03$ \(\textup{M}_J\)), and in an estimate of the system inclination (72.5$^{\circ}$<$i$<84.79$^{\circ}$), with the upper limit set by the non detection of the planetary transit in previous photometric monitoring \citep{Ge2006}.

	In this paper we have not only demonstrated that GIANO is effective at studying the atmosphere of non-transiting planets, but thanks to its wide spectral coverage we have also attained the first detection of methane in a hot Jupiter with the HRS technique. In past studies, methane had been sought in the atmospheres of exoplanets using both low and high spectral resolution data, but without a clear success.\footnote{Low spectral resolution HST NICMOS data had been interpreted in terms of a detection of methane in the atmosphere of HD189733b \citep{Swain2008}, only to be subsequently understood in terms of  systematic effects and unmodeled residuals \citep{Gibson2011}.} These difficulties can be possibly reconciled with the detection presented here if we consider that low spectral resolution space-based data obtained with HST WFC3 do not extend longward of 1.7 $\mu m$, not reaching the K-band and part of the H-band where the strongest methane lines are found. Instead, at high spectral resolution CRIRES pre-upgrade and NIRSpec do have wide spectral coverage respectively of (1.0-5.0) $\mu m$ and (0.95-5.4) $\mu m$, but during one observation only a much smaller spectral region than that covered with GIANO is effectively used.

	In addition to spectral coverage and resolution, there are two further aspects that could explain our success in detecting methane. The first one is the differences in the fine detail of the data analysis that we have used, the second one is the planet's chemistry. However, it is difficult at the moment, on the basis of the few observed comparison objects, to determine which element plays the most important role. A greater number of observed planets could help to clarify this aspect. However, what we can conclude is that our CH$_4$ detection cannot be explained by advances in the line-list quality. Indeed, we use a line-list originally compiled for low-temperature applications, that certainly does not provide a decisive improvement (in terms of e.g. line-list inaccuracies) with respect to those used in previous analysis that ended up in no detections. That being said, our line-list must surely be improved. Indeed, having been built for low-temperature applications, it is likely to miss spectral lines too weak to be relevant at such temperatures, but arguably important at the high temperatures of hot Jupiters. Recently, high-temperature methane line-lists have become available (e.g. \citealt{Yurchenko14}) and we are planning to expand our modeling framework to incorporate them in future work.
	
	The presence of methane in the atmosphere of a hot Jupiter is not so unexpected if we look at theoretical prediction papers (e.g. \citealt{Madhusudhan2012, Madhusudhan2016,Moses2011,Drummond2018, Tsai2018}), although there still is a variety of possible predictions on the expected methane VMR. Our results appear consistent with recent studies conducted on atmospheres similar to that of HD 102195b by \citet{Drummond2018, Tsai2018}.

	As we said the lack of coverage of some spectral regions could be a problem for some spectrographs in the search for methane. 
	Unfortunately, even for GIANO, it can undermine the research of molecules, in particular carbon monoxide. Indeed, we argue that the non detection of this molecule in the GIANO spectra is primarily due to the fact that CO absorbs only partially in the H and K bands and has a lower density of spectral lines.
	
	The results presented here constitute a new demonstration of the efficacy of GIANO for the purpose of atmospheric characterization measurements, but they come with an important caveat. Overall, the useful spectral range is somewhat reduced with respect to that used in the \citet{Brogi2018} study, with fewer orders successfully calibrated in wavelength and a larger number of discarded orders in the cross-correlation analysis. These differences, as far as the analysis of pre-upgrade GIANO observations are concerned, can be understood once the following considerations are taken into account. First, the investigation carried out in this work is based on an independently developed analysis pipeline, that is similar in philosophy to that used in \citet{Brogi2018}, but differs in the fine structure details. Second, compared to e.g. CARMENES and SPIRou, GIANO in its fiber-fed configuration  has a low instrumental stability, which prevents us from using offline calibration frames (e.g. lamp spectra taken before / after the observations) to obtain an acccurate absolute wavelength solution across all orders. GIANO data rely instead on simultaneous calibration using telluric and (where possible) stellar lines. This means that the effectively usable spectral range can be a function of the night conditions (e.g. variable humidity affects the strength of telluric lines), of the achieved S/N per spectrum (that impacts the quality of line centroid determination particularly in the spectral regions where modal noise is strongly present), and of the spectral type of the exoplanet host star. As mentioned in \citet{Brogi2018}, important improvements are expected from the recent GIANO upgrade to a slit-fed instrument (GIANO-B) at the Nasmyth B focus of the TNG, which can be coupled to HARPS-N in the GIARPS simultaneous observing mode \citep{Claudi2017}.
	
	Finally, our analysis can have important consequences for the future characterization of exoplanetary atmospheres. In fact, HD 102195 being $1-3$ $K$-band magnitudes fainter compared to non-transiting systems studied until now with 8-meter telescopes, our study opens the doors to atmospheric characterization measurements with 4-m class telescopes for a an important sample of exoplanets.
	We are now entering a new era for exoplanetary atmospheres characterization with near-infrared, high-resolution spectrographs. 
		CARMENES has started to give its first results \citep{Alonso2019}, GIANO-B has recently come online, and SPIROU,  NIRPS, and CRIRES+ will be available soon. These instruments are extraordinary both in terms of spectral range and throughput, and will be very well suited to perform studies of exoplanetary atmospheres such as the one presented here. The systematic measurements of molecular compounds at high spectral resolution will permit us to create an ''atlas'' of strongly irradiated exoplanetary atmospheres, and in the future, with the coming of the ELTs, it will be possible to apply the same HRS technique to search for biomarkers in the atmospheres of Earth twins \citep{Snellen2013, Rodler2014}.

	\begin{acknowledgements} We wish to thank an anonymous referee for a careful reading and many useful suggestions that helped to improve an earlier version of the manuscript. We thank F. Borsa, I. Carleo, M. Damasso, R. de Kok, A. Maggio, I. Pagano for lending initial impetus to this analysis. G.G. acknowledges the financial support of the 2017 PhD fellowship programme of INAF. P.G. gratefully acknowledges support from the Italian Space Agency (ASI) under contract 2018-24-HH.0. This work has made use of data from the European Space Agency (ESA) mission {\it Gaia} (\url{https://www.cosmos.esa.int/gaia}), processed by the {\it Gaia} Data Processing and Analysis Consortium (DPAC, \url{https://www.cosmos.esa.int/web/gaia/dpac/consortium}). Funding for the DPAC has been provided by national institutions, in particular the institutions participating in the {\it Gaia} Multilateral Agreement. \end{acknowledgements}		
	\bibliographystyle{aa}
	\bibliography{ref}
	
\end{document}